\documentclass[aps,prl,preprint,superscriptaddress]{revtex4-1}

\usepackage{amsfonts}
\usepackage{hyperref}
\usepackage[usenames,dvipsnames]{color}

\usepackage{amssymb,amsmath,graphicx}

\begin{document}


\title{Robust single-shot spin measurement with 99.5\% fidelity in a quantum dot array}


\author{Takashi Nakajima}
\email[Corresponding authors: ]{nakajima.physics@icloud.com}
\author{Matthieu R. Delbecq}
\affiliation{Center for Emergent Matter Science, RIKEN, 2-1 Hirosawa, Wako-shi, Saitama 351-0198, Japan}
\author{Tomohiro Otsuka}
\affiliation{Center for Emergent Matter Science, RIKEN, 2-1 Hirosawa, Wako-shi, Saitama 351-0198, Japan}
\affiliation{JST, PRESTO, 4-1-8 Honcho, Kawaguchi, Saitama, 332-0012, Japan}

\author{Peter Stano}
\author{Shinichi Amaha}
\author{Jun Yoneda}
\affiliation{Center for Emergent Matter Science, RIKEN, 2-1 Hirosawa, Wako-shi, Saitama 351-0198, Japan}

\author{Akito Noiri}
\author{Kento Kawasaki}
\affiliation{Department of Applied Physics, University of Tokyo, 7-3-1 Hongo, Bunkyo-ku, Tokyo 113-8656, Japan}

\author{Kenta Takeda}
\author{Giles Allison}
\affiliation{Center for Emergent Matter Science, RIKEN, 2-1 Hirosawa, Wako-shi, Saitama 351-0198, Japan}

\author{Arne Ludwig}
\author{Andreas D. Wieck}
\affiliation{Lehrstuhl f\"{u}r Angewandte Festk\"{o}rperphysik, Ruhr-Universit\"{a}t Bochum, D-44780 Bochum, Germany}

\author{Daniel Loss}
\affiliation{Department of Physics, University of Basel, Klingelbergstrasse 82, 4056 Basel, Switzerland}
\affiliation{Center for Emergent Matter Science, RIKEN, 2-1 Hirosawa, Wako-shi, Saitama 351-0198, Japan}

\author{Seigo Tarucha}
\email[]{tarucha@ap.t.u-tokyo.ac.jp}
\affiliation{Department of Applied Physics, University of Tokyo, 7-3-1 Hongo, Bunkyo-ku, Tokyo 113-8656, Japan}
\affiliation{Center for Emergent Matter Science, RIKEN, 2-1 Hirosawa, Wako-shi, Saitama 351-0198, Japan}


\date{\today}

\begin{abstract}
We demonstrate a new method for projective single-shot measurement of two electron spin states (singlet versus triplet) in an array of gate-defined lateral quantum dots in GaAs. The measurement has very high fidelity and is robust with respect to electric and magnetic fluctuations in the environment.
It exploits a long-lived metastable charge state, which increases both the contrast and the duration of the charge signal distinguishing the two measurement outcomes.
This method allows us to evaluate the charge measurement error and the spin-to-charge conversion error separately.
We specify conditions under which this method can be used, and project its general applicability to scalable quantum dot arrays in GaAs or silicon. 
\end{abstract}


\maketitle



%

Improving measurement fidelities of qubits is an important step to progress with quantum technologies. Apart from being one of the basic constituents of quantum computation\cite{divincenzo2000:FF}, or even means to perform it\cite{raussendorf2001:PRL}, precise measurements of qubits are indispensable for error correction protocols\cite{shor1995:PRA, kitaev2003:AP, Fowler2009}, or any feedback method in general \cite{griffiths1996:PRL}. Suppressing measurement errors also boosts sensitivity and time resolution of sensors\cite{armen2002:PRL, taylor2008:NP} and, by allowing the manipulations to be performed with less averaging and thus faster, can directly enhance the qubit quality factor\cite{shulman2014:NC, delbecq2016:PRL}.

\begin{figure}
\includegraphics[width=0.7\columnwidth]{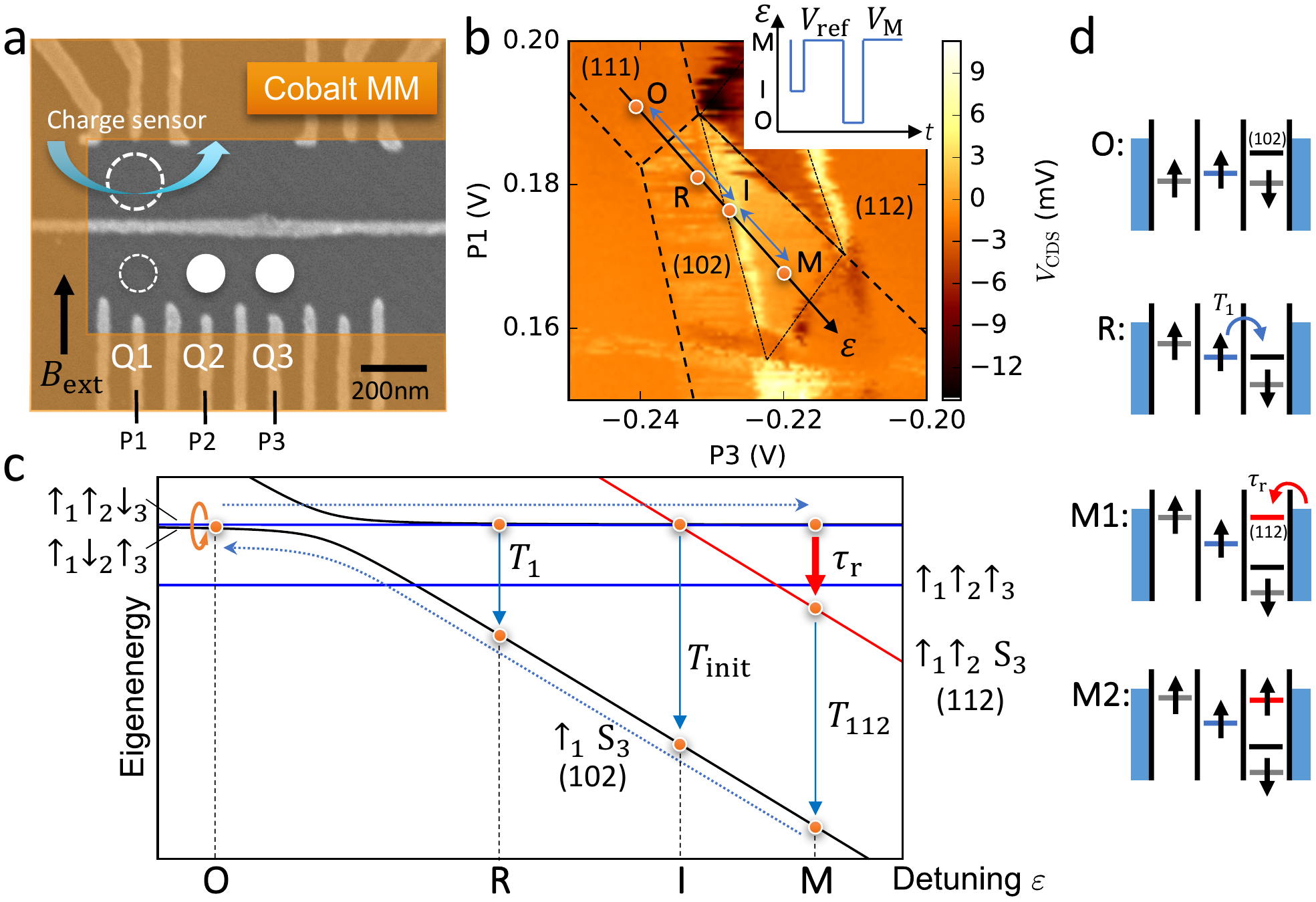}%
\caption{\label{fig1} (a) SEM micrograph of a device similar to the one measured. An array of quantum dots is fabricated with the proximal charge sensor and a top cobalt micromagnet layer (orange-shaded area). The dot Q1 (dashed white circle) is idle and singly occupied throughout the measurement, while the spin pair in Q2 and Q3 (filled white circles) is measured. The fourth dot is not used. (b) Charge stability diagram around the (111)-(102) transition, taken with the application of the I/O/M pulse cycles shown in the inset. The positions of the initialization (I), operation (O), and measurement (M, R) configurations are denoted by circles along with the detuning axis $\varepsilon$. The main panel shows $V_{\text{CDS}}=V_\text{M}-V_\text{ref}$, the difference of the charge sensor signal before ($V_\text{ref}$) and after ($V_\text{M}$) pulsing to O to cancel out the slow drift and the smooth landscape of the background signal. (c) Energy spectrum (solid lines) as a function of $\varepsilon$ for the states labeled according to their charge and spin. The dotted arrows show transitions upon gate pulses and the system can make relaxations (solid arrows) with the corresponding rates labeled during each step. (d) Energy configurations in each pulse step.
}
\end{figure}

For spin qubits in gate-defined quantum dots, which are among prime candidates to realize scalable qubits in solid state \cite{loss1999:PRA, kloeffel2013:ARCMP,Taylor2005}, the first single-shot measurements of a spin-half qubit exploited the spin-dependent energy and tunnel rate and reached fidelities around 80-90\%\cite{elzerman2004:N, hanson2005:PRL}. Later, the development of the rf-reflectometry technique\cite{reilly2007:APL} permitted to use the Pauli spin blockade \cite{ono2002:S,petta2005:S} for a single-shot measurement of a singlet-triplet qubit in double quantum dots with 90\% fidelity\cite{barthel2009:PRL}.
This was further advanced by optimizing the charge sensor sensitivity\cite{barthel2010:PRB} up to the recent value of 98\% reported in Ref.~\cite{shulman2014:NC}. 

Despite the impressive progress, quantum dot spin qubits have been falling short in this respect to other systems, most notably those based on nuclear spins of impurities accessed electrically \cite{pla2013:N} or optically \cite{waldherr2014:N}. To further increase fidelity is not easy, as the signal-to-noise ratio of the charge sensor is limited by the electrical noise in the measurement circuitry and the short lifetime of the spin-blockade state. The latter issue becomes even more serious in the presence of a micromagnet-induced field gradient, which is necessary for fast \cite{yoneda2015:APE} and addressable \cite{Pioro-Ladriere:2008kx} spin manipulations. More importantly, the lifetime of the state being detected is sensitive to both electric and magnetic disturbances, which can drastically degrade the measurement fidelity \cite{barthel2012:PRB}.

Here we implement a single-shot measurement distinguishing two-electron spin states (singlet $\text{S}$ versus spin-unpolarized triplet $\text{T}_{0}$) inside a quantum dot array with 99.5\% fidelity.
It relies on the Pauli spin blockade, but using a different spin to charge conversion, first identified in Ref.~\cite{studenikin2012:APL}. We find that the method leads to a substantial fidelity boost and is robust with respect to environmental fluctuations, both magnetic and electric. This demonstrates that electronic spin qubits can reach measurement fidelities comparable to the highest achieved in solid state, and above the threshold for fault-tolerant quantum computing\cite{Fowler2009,Martinis2015}, without sacrificing their essential advantages of speed \cite{yoneda2014:PRL} and scalability \cite{Noiri2016, otsuka2016:SR}. Furthermore, the high-fidelity measurement allows us to unravel the underlying mechanism of the spin-to-charge conversion error, which is generally present but has been obscured in spin-blockade measurement.

The device is a gate-defined array of quantum dots fabricated on a GaAs/AlGaAs heterostructure with a charge sensor and a cobalt micromagnet on the top, as shown in Fig.~1(a). It was placed in a dilution refrigerator with a base temperature of $\sim 20\,\text{mK}$ and an in-plane magnetic field of $B_\textrm{ext}=0.7\,\text{T}$ was applied. The three left-most dots (Q1-Q3) are used, while the gate electrodes for the fourth dot are grounded. The dot Q1 is kept singly occupied and decoupled from the rest by a high tunneling barrier throughout the experiment. Figure 1(b) shows the relevant part of the charge stability diagram taken as a function of DC gate voltages on P1 and P3, with applying the voltage pulse cycles of the shape shown in the inset. The high-fidelity measurement of the spin states is realized in the bright triangular region shown in Fig.~1b by using the metastable (112) charge state, as discussed later in detail. The energy spectrum along the black line with an arrow (parameterized by $\varepsilon$, the detuning energy between Q1 and Q3) is given in Fig.~1(c).

The standard single-shot measurement based on the spin blockade works as follows.
A gate voltage pulse is applied to alternate between (102) and (111) charge configurations, denoted in Fig.~1(b) and (c) by R and O, respectively. Waiting in R (reset), the system relaxes into its ground state 
$|\sigma_1 \text{S}_{3}\rangle$, meaning the first dot contains an electron with spin $\sigma_1$, the second dot is empty and the third dot is occupied by the singlet.
Pulsing from here to O (operation), the system starts singlet ($\text{S}$)-triplet ($\text{T}_{0}$) precession. Here $\text{S}$ and $\text{T}_{0}$ are coherent superpositions of two system eigenstates, $\left|\sigma_{1}\uparrow_{2}\downarrow_{3}\right>$ and $\left|\sigma_{1}\downarrow_{2}\uparrow_{3}\right>$, split in energy according to the gradient of the micromagnet field $\Delta B_{23}$. The precession, represented in Fig.1(c) by an orange circular arrow, is ended by pulsing back to R. It converts the spin to charge information, because $\text{S}$ goes adiabatically (in the nanosecond ramp time determined by the circuit bandwidth) over to (102), while $\text{T}_{0}$ will remain in (111) until it decays to (102) by a nearest-neighbor hopping. Since the latter transition requires a change of the spin, it is slow enough (typically microseconds) such that the charge sensor can distinguish (111) from (102) and thus the single-shot spin measurement is accomplished.

The measured histogram of the charge sensor signal integrated over time $t_\text{M}=4\,\mu\text{s}$ after pulsing into R is plotted in Fig~2(a). It is well fitted by assuming that it originates from two discrete values $V_\textrm{T}$ and $V_\textrm{S}$, assigned to (111) and (102), smeared by the Gaussian noise of the charge sensor \cite{barthel2009:PRL}. While integrating the signal longer averages out the noise, it also leads to an overall shift of the signal towards $V_\textrm{S}$, because of the finite lifetime of (111), $T_1$. The latter can be found from the time dependence of the mean value of the sensor signal. This is plotted in Fig.~2(c), and an exponential fit gives $T_1 \approx 9\,\mu\text{s}$.
Therefore, there is an optimal integration time $t_\text{M}$ and a threshold voltage $V_{\text{th}}$ which maximizes the contrast by minimizing the overlap of the two Gaussian-like distributions forming the histogram. This overlap is the infidelity (one minus the fidelity) of the specific {\it charge} measurement, being the measure of the reliability with which one can discriminate the system being initially in (111) versus (102). For the data plotted in Fig.~2(a) the fidelity is $83.8\pm0.8$\%.

Once the (102)/(111) charge state is identified, it is interpreted as the spin singlet/triplet measurement outcome. However, we stress that the fidelities of the charge and spin measurement are not identical, since the spin measurement fidelity is further diminished by the fidelity of the spin to charge conversion.
Our measurement scheme explained below significantly improves the charge measurement fidelity. Its robustness against the magnetic and electric fluctuations also allows us to separately analyze the spin measurement fidelity as discussed below.

The high-fidelity single-shot measurement is performed at the readout point M inside the bright triangular region shown in Fig.~1(b), by taking advantage of the presence of an excited, additionally charged, state (112). As shown in Fig.~1(c), the energy of this state is below that of (111) at M. Upon applying the measurement pulse from O, the singlet goes over to (102) as before, but the charge sequence for a triplet changes. It first loads (with a fast rate $\tau_{r}^{-1}$) an additional electron from the lead into Q3, going to (112), before relaxing (with a slow rate $T_{112}^{-1}$) to the system ground state (102). This has two decisive advantages. First, the charge states to be distinguished differ by the total number of electrons in the system [(102) versus (112)] and not just by their position [(102) versus (111)], which gives a larger signal contrast. Second, the lifetime of the metastable state is longer, which 
diminishes the shift of the triplet signal due to the relaxation in a given integration time\cite{footnote2}.

\begin{figure}
\includegraphics[width=1\columnwidth]{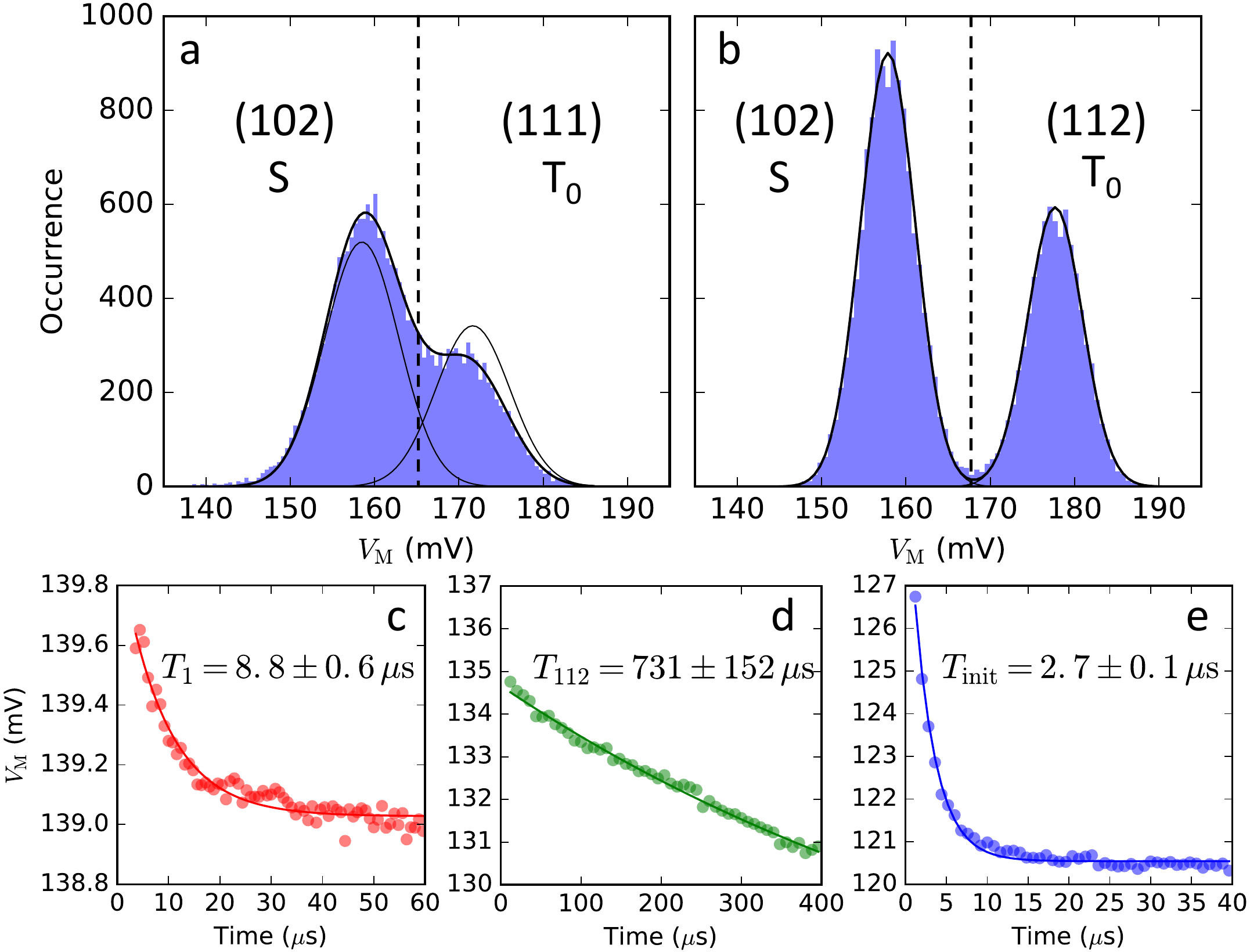}
\caption{\label{fig2}
(a-b) Histograms of the single-shot signals in configuration (a) R ((111) readout; $V_\text{S}=158.5\,\text{mV}$, $V_\text{T}=171.7\,\text{mV}$) and (b) M ((112) readout; $V_\text{S}=157.9\,\text{mV}$, $V_\text{T}=177.7\,\text{mV}$). Thin solid lines show Gaussian distributions for $\text{S}$ and $\text{T}_{0}$ that would have been observed if no relaxation occurred. Dashed vertical lines show the threshold voltages. (c-e) Waiting time dependence of the mean signal $V_\text{M}$ at (c) R (d) M, and (e) I. The values of $V_\text{M}$ differ from those in (a-b) due to the change of the charge sensor condition. The curves are fits to $A+B \exp(-t/T)$ with the corresponding relaxation time $T$ given in each panel.}
\end{figure}

The resulting improvement is clearly visible from the histogram in Fig.~2(b), the analog of Fig.~2(a), with fidelity $99.68 \pm 0.06\%$. The fidelity boost is possible because of the hierarchy of the relaxation times, $T_{112} \gg T_1 \gg \tau_{r}$, which is easily realized in quantum dot arrays. The lifetime of (112), $T_{112}$, is large [$0.7\,\text{ms}$ here, see Fig.~2(d)], because the relaxation from (112) to (102) requires to remove an electron from a dot without a direct access to a lead, a next nearest-neighbor tunneling.
On the other hand, the relaxation of the triplet in (111) to the singlet in (102) is caused by a nearest-neighbor tunneling accompanied by spin mixing, resulting in $T_1 \approx 9\,\mu\text{s}$ [see Fig.~2(c)].
Finally, since the loading of an extra electron from a lead to Q3 is blocked by neither spin nor charge, $\tau_r$ is the smallest. The value of $\tau_{r}^{-1}$ is well above the measurement bandwidth of the charge sensor such that $\tau_{r}^{-1}\gg 10\,\text{MHz}$, and we expect $\tau_{r}^{-1}$ to be equal to the Q3-lead tunnel rate, which can easily reach $100\,\text{MHz}$\cite{House2015}.

The infidelity of the charge measurement is of the order of the small ratios $\tau_r/T_1$ and $t_\text{M}/T_{112}$. Since we 
estimate $\tau_r/T_1 < 10^{-3}$ and $t_\text{M}/T_{112}\approx 5\times 10^{-3}$, the latter dominates the infidelity in the present setup. Employing the theory developed in Ref.~\cite{barthel2009:PRL} allows---from a fit to the histogram---to both evaluate the fidelity and optimize it by choosing the proper integration time $t_\text{M}$ and the threshold voltage [denoted by vertical dashed lines in Fig.~2(a),(b)] used to assign the binary result. This is how we arrived at the value 99.68\%, and dependence of the maximal fidelity on the metastable state lifetime is further illustrated in the Supplemental Material (SM)\cite{SM}. More importantly, the condition 
$\tau_r/T_1 \ll 1$ makes the measurement fidelity insensitive to 
modest variations of $T_1$ 
due to fluctuations of the Overhauser field and electrostatic potential\cite{barthel2012:PRB}. 
This insensitivity to $T_1$ makes our measurement robust throughout a long-term experimental run, which is a major advantage.

Despite the long lifetime of (112), one can perform the spin initialization by inserting an additional pulse step positioned at I in Fig.~1(b). This takes advantage of the increased efficiency of the relaxation at the degeneracy of (112) with (111)\cite{Johnson:2005jw}, which is visible as the bright line (larger signal) along the edge of the triangular readout region in Fig.~1(b).
The corresponding relaxation time is fitted to $T_\text{init} \approx 3$ $\mu$s from the data shown in Fig.~2e, being more than three times smaller than $T_1$.

To evaluate the fidelity of \emph{spin} measurement, however, one has to consider additional errors arising in its conversion to charge by pulsing from O to M. The high-fidelity charge measurement developed here allows us to study this effect separately.
The dominant source of errors is the deviation from the pulse being 
perfectly adiabatic with respect to (111) and (102) singlet-singlet anticrossing. Using the Landau-Zener formula, the probability to move through a state crossing non-adiabatically would give this error as
\begin{equation}
p_\text{n} \sim \exp\left(-\frac{2\pi t_c^2}{\hbar \Delta \epsilon} \Delta t\right),
\label{LZ}
\end{equation}
where $2t_c$ is the energy splitting at the anticrossing, 
and $\Delta \epsilon$ is the change of the energy difference of the crossing states during the pulse time $\Delta t$.
Additional errors, such as photon-assisted charging, spin decay by co-tunneling, or spin relaxation by phonon-emission are, first, not specific to the measurement pulse, and, second, we find these negligible compared to $p_\text{n}$ based on estimates given in SM\cite{SM}.

Instead of estimating $p_\text{n}$ from Eq.~\eqref{LZ}, we directly measure it. To this end, 
we set up a rate-equation model (see section II of SM\cite{SM}) for the previously described I$\to$O$\to$M cycle and derive 
\begin{equation}
P_\text{S}( t ) = a + \frac{v}{2}\, e^{-\left(t/T_2^*\right)^2} \cos( \omega t + \phi) + c \, e^{ -\Gamma t},
\label{fit}
\end{equation}
as the probability to measure signal `S' after the $\text{S}$-$\text{T}_{0}$ precession with an angular frequency $\omega$ for a duration $t$, with $\phi$ an additional phase shift and $T_{2}^{*}$ the ensemble dephasing time. The idea is that the same non-adiabaticity as the one causing the error in the spin to charge conversion, $p_\text{n}$, results in errorneous initialization to the excited (102) state rather than the (111) singlet state at O [see Fig.~3(a)]. If this excited state lifetime $1/\Gamma$ is relatively long, as is the case here, the imperfect initialization is directly visible as an exponentially decaying signal downshift by $c\propto p_\text{n}$, described by the last term in Eq.~\eqref{fit}.

\begin{figure}
\includegraphics[width=1\columnwidth]{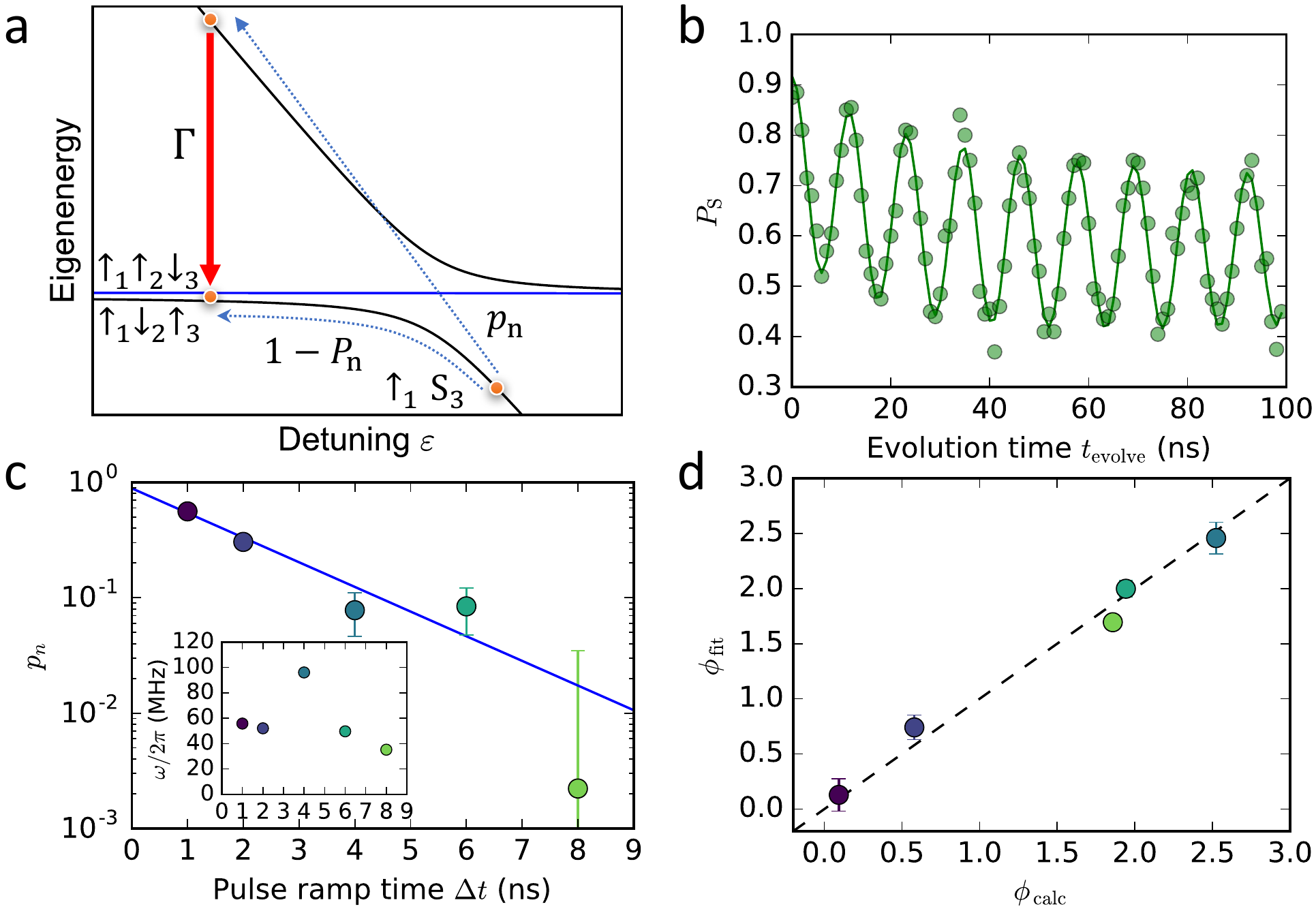}%
\caption{\label{fig3} (a) Schematics of the transition through the singlet $\text{S}$-$\text{S}_3$ (black lines) anticrossing. The blue horizontal line is the energy of the (111) triplet $\text{T}_0$. Notation similar to Fig.~\ref{fig1} is used. (b) Signal oscillation observed in the I$\to$O$\to$M cycle. (c) The probability $p_\text{n}$ as a function of the pulse ramp time $\Delta t$. A solid line is the fit according to Eq.~\eqref{LZ} written in form $\exp(-t/t_0)$, which gives $t_0=1.16\,\text{ns}$. The inset shows the random fluctuation of $\omega$ in the course of measurement. (d) The values of the phase shift $\phi_\text{fit}$ obtained from the fit with Eq.~\eqref{fit} versus the values $\phi_\text{calc}$ calculated from $p_{n}$, $\Gamma$, $\omega$, and $\Delta t$ using the theoretical model (see SM\cite{SM}).}
\end{figure}

Before discussing the other terms of Eq.~\eqref{fit}, Fig.~3(b) shows an exemplary data set, together with the fit according to Eq.~\eqref{fit}. The downshift of the oscillating signal with $t$ is apparent and allows us to extract $p_\text{n}$ and $\Gamma$.
The fit results in $\Gamma=13.8\pm 4.5\,\text{MHz}$ which complies very well with a microscopic model of the quantum dot (see SM).
As shown in Fig.~3(c), we find that $p_\text{n}$ is suppressed substantially by slowing down the pulse ramp between O and M. The observed dependence on the pulse ramp time $\Delta t$ follows the 
scaling suggested by Eq.~\eqref{LZ}. By suppressing the non-adiabaticity error to the value fitted for $\Delta t = 8\,\text{ns}$ to be $p_\text{n} \approx 2 \times 10^{-3}$, we arrive at the spin measurement fidelity of $99.5\%$, with the $0.2\%$ error of the spin to charge transfer\cite{footnote4} and $0.3\%$ error of the charge readout. This constitutes our main result.

We now turn to the remaining parameters of Eq.~\eqref{fit}. Figure 3(d) shows the fitted phase shift $\phi$. We find that it is dominated by the phase acquired during the pulse ramp time of $\Delta t$, rather than by a contribution from $p_\text{n}$, and therefore does not allow us to independently estimate $p_\text{n}$.
Similarly, we find that the values of the offset $a$ and the visibility $v$ are much more susceptible to noise and therefore not reliable to estimate other parameters involved in the model, especially the initialization fidelities into various possible states during waiting at I (see SM).
We believe this is because of the Overhauser field fluctuations. We take them partially into account in Eq.~\eqref{fit} by introducing the dephasing time $T_2^*$, appropriate for weak Gaussian noise in $\omega$. However, short acquisition times which we employ to prolong $T_2^*$\cite{delbecq2016:PRL}, at the same time lead to these fluctuations varying non-uniformly over different, or even during, measurements. These fluctuations are not weak, as we estimate that
the magnetic field gradient can be sometimes as small as the exchange coupling due to the fluctuations. This leads to changes of the precession axis direction and additional initialization errors\cite{Shulman2012}, which our model resulting in Eq.~\eqref{fit} does not take into account.

We would like to make several comments now. First, metastable states such as the one used here are a typical feature found in quantum dot arrays. Second, the presented method is applicable to larger arrays without extensive tuning of the tunnel rates.
Third, we stress that the measurement fidelity is stable with respect to the variation of the Overhauser field, which here leads to variations of the precession frequency.
Despite the variation of $\omega/2\pi$ in a wide range of $35$-$95\,\text{MHz}$ in the course of the measurement as shown in the inset of Fig.~3(c), we did not find any apparent effects on the histogram in Fig.~2(b).
Third, the very long lifetime of the metastable state would enable sequential readout of many spins using a switch matrix and a single transmission line\cite{Hornibrook2015}, which will be an important technical simplification of the circuitry for large-scale quantum computing. 
With spin measurement fidelities achieved here, we estimate that $19$ qubits can be read out with the fidelity above $90\%$\cite{SM}. 
Finally, we suppose that it will be possible to increase the measurement fidelity much further by tuning the dot parameters, especially the dot-dot and dot-lead tunnel rates, that are not extensively optimized in this work.

Before concluding, let us discuss the results presented here from a broader view. Even though we believe that the achieved high fidelity characterizes the measurement of the spin (and not just a charge), it cannot be strictly proven unless the fidelities in other parts of the experiment---spin initialization and manipulation---are higher than, or at least comparable to, the measurement fidelity\cite{footnote1}. The whole cycle as we do here is aimed at observing the $\text{S}$-$\text{T}_{0}$ oscillations. The qubit initialization, coherent rotation and measurement, taken all together can be regarded as a quantum algorithm, perhaps the most simple one. The overall precision of this specific algorithm is revealed by the visibility of the oscillations, to which imperfections of all parts contribute.
Interestingly, we observe a non-monotonic change of the visibility $v$ upon suppressing the measurement errors (see SM), suggesting that fidelities of these other parts are influenced upon changing the pulse time \cite{footnote3}.
Nevertheless, a precise measurement is the first requirement for being able to characterize and confirm the suppression of these imperfections, for which many methods have been suggested. 

In conclusion, we reached 99.5\% fidelity of the single-shot spin measurement in a quantum dot array using a metastable state for the charge readout. It has two advantages, a stronger and a longer lived charge signal corresponding to the two possible measurement results. Requirements for using this method are simple, and we therefore find it generally suited for scalable structures of gate-defined quantum dots in GaAs as well as Si. The high-fidelity measurement will bring the spin qubit platform closer to the error-correction threshold and serve as a useful tool for distant quantum communications in which projection measurement onto a `Bell basis' is essential.

\begin{acknowledgments}
We thank the Microwave Research Group in Caltech for technical support. 
This work is financially supported by CREST, JST (JPMJCR15N2, JPMJCR1675), the ImPACT Program of Council for Science, Technology and Innovation (Cabinet Office, Government of Japan).
TN, TO and JY acknowledge financial support from RIKEN Incentive Research Projects.
PS acknowledges financial support from JSPS KAKENHI Grant Number 16K05411.
TO acknowledges financial support from PRESTO, JST (JPMJPR16N3), JSPS KAKENHI Grant Numbers 25800173 and 16H00817, Strategic Information and Communications R\&D Promotion Programme, Yazaki Memorial Foundation for Science and Technology Research Grant, Japan Prize Foundation Research Grant, Advanced Technology Institute Research Grant, the Murata Science Foundation Research Grant, Izumi Science and Technology Foundation Research Grant, TEPCO Memorial Foundation Research Grant, The Thermal \& Electric Energy Technology Foundation Research Grant.
AN acknowledges support from Advanced Leading Graduate Course for Photon Science (ALPS).
ST acknowledges financial support by JSPS KAKENHI Grant Numbers 26220710 and JP16H02204.
AL and ADW acknowledge gratefully support of Mercur Pr-2013-0001, DFG-TRR160, BMBF – Q.com-H 16KIS0109, and the DFH/UFA CDFA-05-06.
\end{acknowledgments}


\begin{thebibliography}{99}
\bibitem{footnote2}
Alternatively, one could integrate signal longer to suppress the noise, though this kind of optimization was not done here.
\bibitem{footnote4}
The spin to charge conversion error is most probably even smaller than this value, see the discussion around Eq.~(19) in SM\cite{SM}.
\bibitem{footnote1}
Their characterization in turn requires a precise measurement, so that non-trivial bootstrap methods will be necessary in confirming high fidelities of a quantum algorithm.
\bibitem{footnote3}
This effect is currently under investigations.
\end{thebibliography}
\end{document}